\documentclass[%
 reprint,
 superscriptaddress,
nofootinbib,
 amsmath,amssymb,
 aps, physrev,
]{revtex4-2}

\usepackage{graphicx}% Include figure files
\usepackage{dcolumn}% Align table columns on decimal point
\usepackage{bm}% bold math
\usepackage{multirow}
\usepackage{amsmath,booktabs,siunitx} 
\usepackage[dvipsnames]{xcolor}
\begin{document}

\preprint{APS/123-QED}

\title{The role of externally-driven density fluctuations in the enhancement of beam spray
%Enhanced beam spray seeded by speckle-driven density fluctuations
%Enhancement of laser beam spray by separate speckled laser beams
}% Force line breaks with \\

\author{K.R. McMillen}
 \email{Contact Author: kmcm@lle.rochester.edu}
 \affiliation{Laboratory for Laser Energetics, 250 East River Road, Rochester, New York 14623, USA}
 \affiliation{Department of Physics and Astronomy, University of Rochester, Rochester, New York 14623, USA}

\author{I.D. Min-Roberts}
 \affiliation{Department of Physics, University of Alberta, Edmonton, Alberta T6G 2E1, Canada}
 
\author{R.K. Follett}
 \affiliation{Laboratory for Laser Energetics, 250 East River Road, Rochester, New York 14623, USA}
 
 \author{A.L. Milder}
 \affiliation{Laboratory for Laser Energetics, 250 East River Road, Rochester, New York 14623, USA}
  
\author{J. Katz}
 \affiliation{Laboratory for Laser Energetics, 250 East River Road, Rochester, New York 14623, USA}

\author{D.J. Haberberger}
 \affiliation{Laboratory for Laser Energetics, 250 East River Road, Rochester, New York 14623, USA}
 
\author{J.R. Hernandez Gordillo}
 \affiliation{Laboratory for Laser Energetics, 250 East River Road, Rochester, New York 14623, USA}

\author{P.A. Michel}
 \affiliation{Lawrence Livermore National Laboratory, 7000 East Avenue, Livermore, California, 94550, USA}

\author{H.G. Rinderknecht}
 \affiliation{Laboratory for Laser Energetics, 250 East River Road, Rochester, New York 14623, USA}
 \affiliation{Department of Physics and Astronomy, University of Rochester, Rochester, New York 14623, USA}

\author{D. Turnbull}
 \affiliation{Laboratory for Laser Energetics, 250 East River Road, Rochester, New York 14623, USA}

\author{J.P. Palastro}
 \affiliation{Laboratory for Laser Energetics, 250 East River Road, Rochester, New York 14623, USA}

\author{D.H. Froula}
 \affiliation{Laboratory for Laser Energetics, 250 East River Road, Rochester, New York 14623, USA}
 \affiliation{Department of Physics and Astronomy, University of Rochester, Rochester, New York 14623, USA}

\author{W. Rozmus}
 \affiliation{Department of Physics, University of Alberta, Edmonton, Alberta T6G 2E1, Canada}

\author{J.L. Shaw}
 \email{Contact Author: jshaw05@lle.rochester.edu}
 \affiliation{Laboratory for Laser Energetics, 250 East River Road, Rochester, New York 14623, USA}
 \affiliation{Department of Physics and Astronomy, University of Rochester, Rochester, New York 14623, USA}

\begin{abstract}

%Speckled laser beams, commonly used to ionize and heat plasma for inertial confinement fusion and high-energy density experiments, are shown to drive plasma-density fluctuations that induce beam spray in other laser beams. Experiments show enhanced levels of beam spray of a 1053-nm beam while 351-nm speckled heater beams are on compared to after they have turned off. These observations indicate that the high-intensity speckles sustain an increased level of steady-state density fluctuations. Modeling of the speckle-driven plasma-density fluctuations shows amplitudes that are orders of magnitude larger than the level of thermal fluctuations. Supporting simulations show that these fluctuations seed filamentation in the 1053-nm beam, which leads to the observed levels of beam spray.

Speckled laser beams, commonly used to ionize and heat plasma for inertial confinement fusion and high-energy-density experiments, are shown to drive long-wavelength plasma-density fluctuations that induce beam spray in other laser beams. Experiments show that a 1053 nm beam undergoes increased beam spray when 351 nm speckled heater beams are on. These observations indicate that the high-intensity speckles sustain an increased level of steady-state density fluctuations. Modeling of these fluctuations predicts amplitudes that are orders of magnitude larger than the level of thermal fluctuations. Beam propagation simulations show increased beam spray due to self-focusing and filamentation when heater-driven density fluctuations are included.
\end{abstract}

\maketitle

%\tableofcontents
Understanding the seeding and growth of laser-plasma instabilities (LPI) is essential for isolating and mitigating instabilities for inertial confinement fusion research \cite{labaune_Laser_2000, fuchs_Modification_2000, montgomery_Two_2016}, laser-plasma amplifiers \cite{haberberger_Hot_2021, trines_New_2020, vieux_High_2020, lancia_Experimental_2010, ping_Development_2009}, plasma-based optics \cite{peng_Nonlinear_2019a, glenzer_Symmetric_2010, michel_Dynamic_2014b, turnbull_High_2016}, and Thomson scattering experiments \cite{hansen_Mitigation_2019, milder_Measurements_2021, turnbull_Thomson_2026}. In experiments, instabilities have finite time and space to grow, and so elevated levels of fluctuations can cause otherwise harmless instabilities to become detrimental.

%Part of this understanding requires knowledge about the level of fluctuations present in the plasma that can seed these instabilities and lead to deleterious effects such as fast-particle generation, scattering of laser light, and enhanced energy absorption.
 
Beam smoothing techniques such as spatial smoothing using phase plates \cite{kato_random_1984}, polarization smoothing \cite{boehly_reduction_1999}, and smoothing by spectral dispersion \cite{skupsky_improved_1989} purposefully reduce the coherence properties of the laser light to disrupt the growth of LPI. The phase plates \cite{neauport_Design_2003, marozas_Fourier_2007} used on the National Ignition Facility \cite{haynam_National_2007}, OMEGA 60 \cite{boehly_upgrade_1995}, and Laser Mégajoule \cite{neauport_Laser_2024} introduce phase perturbations across the near field beam profile to control the shape of the far-field focal spot on target. Previous theoretical and computational work has predicted that the speckles created by beams using phase plates can ponderomotively and thermally drive plasma-density perturbations and low-frequency ion-acoustic waves \cite{rozmus_enhanced_1994, senecha_temperature_1998, brantov_plasma_1999, grech_plasma_2006}. These fluctuations can couple to LPI such as forward stimulated Brillouin scattering and filamentation, which can result in spectral and angular broadening, or beam spray, of the speckled beam \cite{schmitt_Timedependent_1998, moody_First_1999, moody_experimental_2000, fuchs_Experimental_2001, malka_Enhanced_2003, grech_Coherent_2009, turnbull_Beam_2022, ruyer_Statistical_2025}. The spectral and angular broadening results in a loss of temporal and spatial coherence that is sometimes referred to as plasma-induced smoothing.

\begin{figure}[b!]
    \centering
    \includegraphics[trim = 10.7cm 0cm 10.7cm 5.1cm, clip, width=1.00\linewidth]{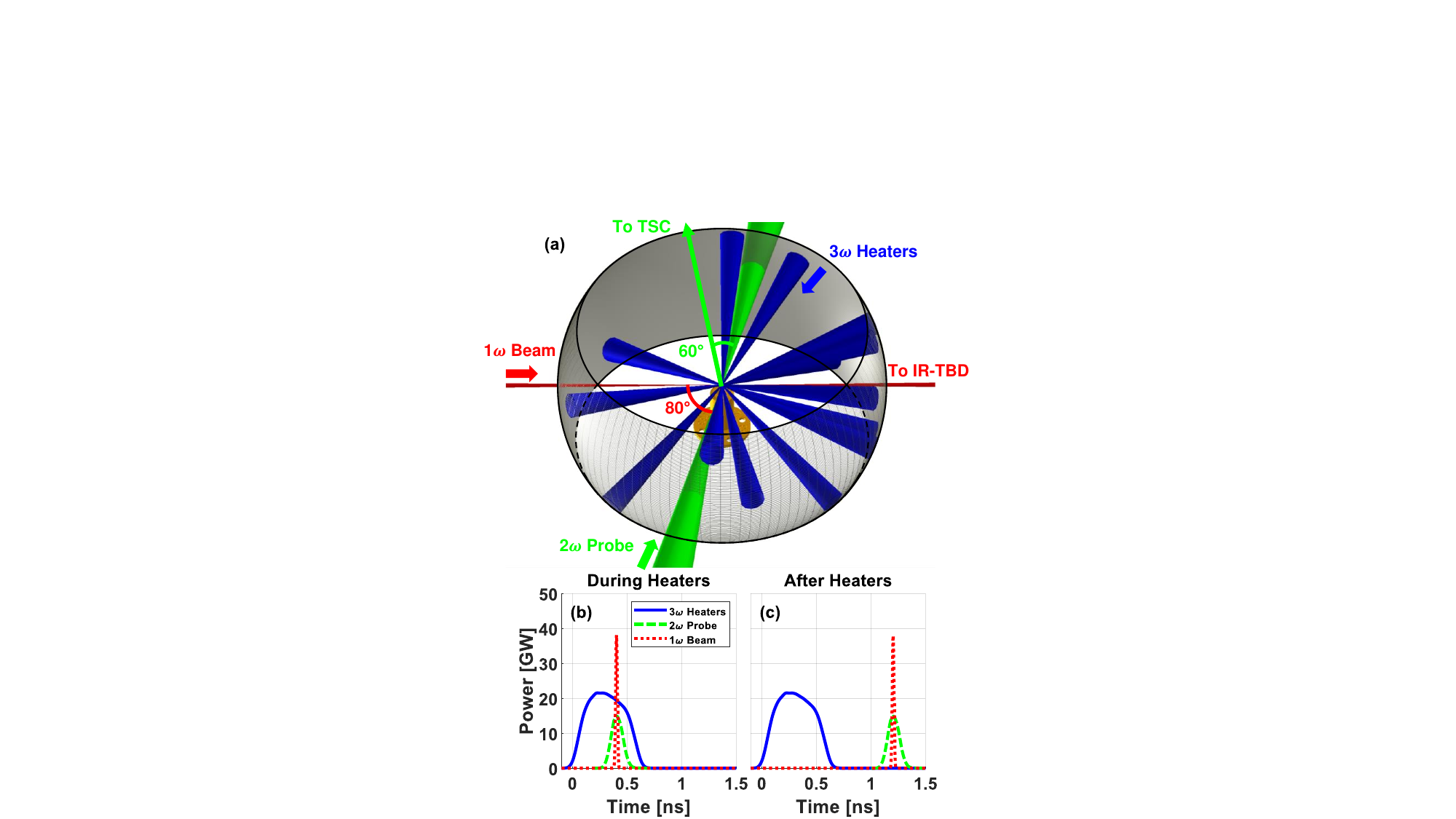}
    \caption{(a) Experimental setup showing the 3$\omega$ heater beams (blue), 2$\omega$ Thomson-scattering probe (green), and 1$\omega$ beam (red) focused from OMEGA 60 port positions above a gas-jet nozzle (gold). The infrared-transmitted beam diagnostic (IR-TBD) images the transmitted 1$\omega$ beam profile, and the Thomson scattering collection (TSC) system collects Thomson-scattered probe light (see appendix for diagnostic details). The timing configurations (b) during heater beams and (c) after heater beams. The power of one heater beam, the probe beam, and the 1$\omega$ beam are shown in the solid blue, dashed green, and dotted red lines, respectively.}
    \label{fig:setup}
\end{figure}
%While plasma-induced smoothing has historically referred to self-induced smoothing of beams, smoothing induced by separate beams has also been observed in experiments. Moody \textit{et al.} \cite{moody_First_1999, moody_experimental_2000} reported small-angle spreading of a spatially smoothed beam by density perturbations driven by intensity modulations in the plasma-forming beams.

In this Letter, we present the role of external speckle-driven density fluctuations in the seeding and enhancement of beam spray. We show that speckled laser beams sustain increased levels of stationary density fluctuations that induce beam spray in separate beams. We measure the beam spray of an unsmoothed beam after propagation through a well-characterized plasma both with and without stationary speckle-driven fluctuations. A new model of these steady-state density perturbations was developed which incorporates nonlocal thermal modifications and realistic beam geometry. This model shows that at an overlapped intensity of $8\times 10^{13}$ W/cm$^2$, steady-state density perturbations can reach many orders of magnitude larger than the zero-frequency thermal fluctuations. Beam propagation simulations show that including these density perturbations produces beam spray consistent with that observed in experiments.

The experimental geometry is shown in Fig. \ref{fig:setup}(a), in which fourteen 351-nm (3$\omega$) heater beams from the OMEGA 60 laser system ionized and heated a gas-jet target. A 1053-nm (1$\omega$) OMEGA EP \cite{meyerhofer_performance_2010} short-pulse beam propagated through the resulting plasma. Simultaneously, a 527-nm (2$\omega$) Thomson-scattering probe measured plasma conditions as a function of distance across the gas-jet plasma. A Mach 5, 2-mm-exit-diameter nozzle positioned 2 mm below the target chamber center (TCC) produced the gas jet using 40--50 psi of N$_2$ \cite{hansen_supersonic_2018, mcmillen_validation_2024}.

Each 3$\omega$ heater beam was focused to TCC using an f/6.7 optic and delivered 10 J/beam over a 500 ps full-width-at-half-maximum (FWHM) duration, resulting in an average overlapped intensity of $8\times10^{13}$ W/cm$^2$. Each heater beam utilized a distributed polarization rotator for polarization smoothing and an SG5 distributed phase plate, which produced a 5th-order super-Gaussian spot with a 660-$\mu$m FWHM. Smoothing by spectral dispersion was not used on any beams so that the 3$\omega$ beams' speckles did not move over time. This stationary behavior allowed the speckles to ponderomotively and thermally drive large-amplitude stationary density perturbations.

A 1$\omega$ OMEGA EP short-pulse beam was used to investigate the effect of speckle-driven plasma-density fluctuations on beam propagation. The beam was apodized to f/50 to minimize the effects of phase aberrations in the beam and to extend the confocal region across the complete $\sim$1-mm-long plasma. This apodization generated a Gaussian vacuum focal spot with a $40$ $\mu$m FWHM. The beam had 0.78 $\pm$ 0.08 J of energy and was stretched to a pulse duration of 30 $\pm$ 1 ps, producing a focal spot intensity of $1.4\pm 0.14\times10^{15}$ W/cm$^2$.

The 1$\omega$ beam and 2$\omega$ probe were delayed 400 or 1200 ps relative to the heater beams, which allowed the 1$\omega$ beam to interact with the plasma during or after the heater beams, respectively. This also allowed the Thomson-scattering probe to measure the relevant plasma conditions. The 400- and 1200-ps timing configurations are shown in Fig. \ref{fig:setup}(b) and (c), respectively.

\begin{figure}[h]
    \centering
    \includegraphics[trim = 10.7cm 0cm 10.7cm 8.6cm, clip, width=1.00\linewidth]{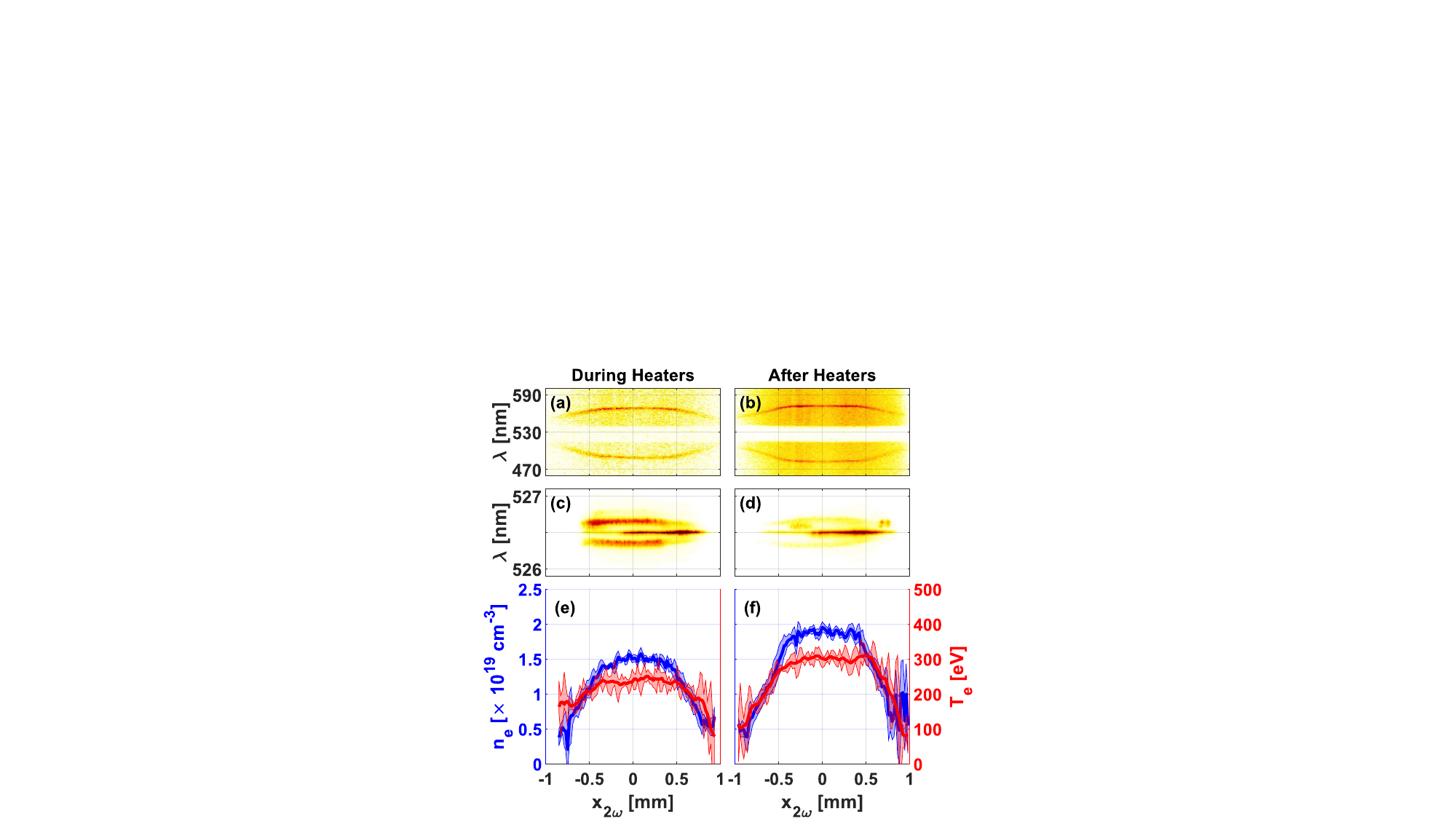}
    \caption{Representative examples of spatially- and spectrally-resolved Thomson-scattered electron and ion features (a, c) during and (b, d) after the heater beams, respectively. (e) and (f) Moving average of the electron density (blue) and electron temperature (red). $x_{2\omega}$ is the distance from TCC along the propagation direction of the 2$\omega$ probe. The shaded regions indicate the 2$\sigma$ standard deviations. %The bright unshifted feature in (c) and (d) corresponds to the specular reflection of $2\omega$ light off of the gas-jet nozzle.
    }
    \label{fig:TSData}
\end{figure}

Plasma conditions were determined from spatially-resolved Thomson scattering data shown in Fig. \ref{fig:TSData}(a)-(d). Due to the symmetric distribution of heater beams about the gas-jet nozzle, the plasma conditions were assumed to be cylindrically symmetric. The electron density and temperature profiles in Fig. \ref{fig:TSData}(e) and (f) were obtained by fitting the measured spectra (a)-(d) using an automatic differentiation-based gradient descent method \cite{milder_qualitative_2024}.
%and averaged over the 20-$\mu$m spatial resolution of the diagnostic. 
In the configuration during (after) the heater beams, the spectra were fit using a calculated super-Gaussian electron distribution function of order 2.86 (2.80) \cite{matte_NonMaxwellian_1988}. The resulting plasma density, ionization, and temperature in the plateau region of the plasma from $x \approx -0.5$ to $0.5$ mm are $n_{e0} = 1.34 \pm 0.05 \ (1.85 \pm 0.05 ) \times 10^{19}$ cm$^{-3}$, $Z = 5\ (5)$, and $T_e = 222 \pm 13\ (299\pm11)$ eV. The difference in plasma conditions resulted primarily from different gas-jet backing pressures. The configuration during the heater beams used 44 psi, while the configuration after the heater beams used 51 psi. %The result of this is a negligible 2.5\% change in the ponderomotive coupling strength (scales as $\sim n_e/T_e$) between the configurations.

\begin{figure}[t]
    \centering
    \includegraphics[trim = 10.7cm 0cm 10.7cm 4cm, clip, width=1.00\linewidth]{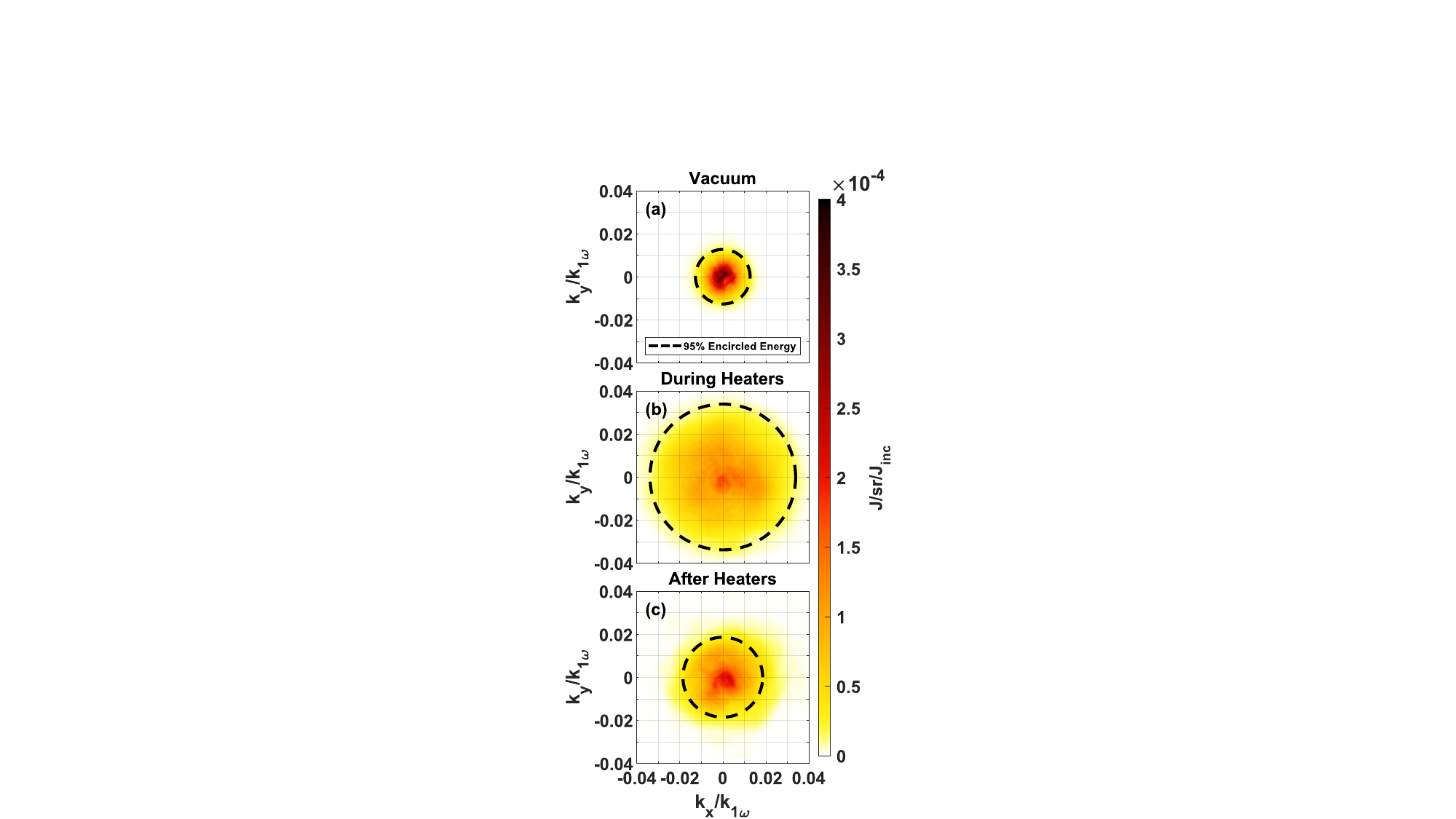}
    \caption{Experimental beam-spray measurements at $1.4\times 10^{15}$ W/cm$^2$ after (a) vacuum propagation, (b) plasma propagation during heater beams, and (c) plasma propagation after heater beams. The dashed line represents the 95$\%$ encircled-energy contour. Data are normalized by their respective incident beam energies.}
    \label{fig:beamspray}
\end{figure}

The angular profiles of the 1$\omega$ beam in Fig. \ref{fig:beamspray} clearly show increased beam spray after propagation through plasma with the speckled heater beams on. Beam spray was quantified by the angle containing 95\% of the IR-TBD-collected energy denoted as $(k_\perp/k_{1\omega})_{95}$ with the transverse wavenumber, $k_\perp^2 = k_x^2 + k_y^2$, and $k_{1\omega} = 2\pi/1.053$ $\mu$m. Compared to the vacuum transmission case shown in Fig. \ref{fig:beamspray}(a) ($(k_\perp/k_{1\omega})_{95} = 12.7$ mrad), propagation during heaters (Fig. \ref{fig:beamspray}(b)) showed a $\sim2.7$ times increase in beam spray out to $(k_\perp/k_{1\omega})_{95} = 33.8$ mrad. This small-angle beam spray corresponds to long-wavelength fluctuations induced by the speckled heater beams via $\theta\approx k_\perp/k_{1\omega}$. Figure \ref{fig:beamspray}(c) shows the case where the 1$\omega$ propagated through after the heaters. Without the speckle-driven perturbations, the fluctuation level is reduced to thermal levels. The decreased fluctuation level reduces the angle of the beam spray to $k_\perp/k_{1\omega} = 18.5$ mrad.

Additional experiments, not shown here, at a 1$\omega$ beam intensity of $3.4\times10^{14}$ W/cm$^2$ demonstrated that the angle of the beam spray during heaters increases with intensity. This indicates beam spray is produced by a nonlinear mechanism such as local self-focusing and filamentation of the 1$\omega$ beam through speckle-driven density perturbations. This distinguishes the severity of the problem from the linear scattering observed by Moody \textit{et al.} \cite{moody_First_1999, moody_experimental_2000}. Note that whole-beam filamentation does not contribute to the beam spray. Although the 1$\omega$ beam has high power, its 30-ps pulse duration is shorter than the timescale on which filamentation develops, $r_{1\omega}/c_s \sim 200$ ps \cite{michel_Introduction_2023}. $r_{1\omega} = 20$ $\mu$m is the half-width-at-half-maximum of the beam and $c_s \approx 1\times 10^5$ m/s is the sound speed. Furthermore, the highest growth-rate wavevector for whole-beam filamentation corresponds to spray at much larger angles than observed, near $k_\perp/k_{1\omega}\approx 0.1$.

The model of the speckle-driven density perturbations, $\delta n_s$, assumed they are in steady state, as the time scale for the speckles to dig their density perturbations is $\lambda_{3\omega}(f_\#)_{eff}/ c_s\approx 3.5$ ps. $\lambda_{3\omega} = 351$ nm and $(f_\#)_{eff} \sim 1$ is the effective f-number corresponding to the highest-intensity speckles produced by many overlapped beams \cite{follett_Independenthotspot_2022}. As a result, the density perturbations were modeled in Fourier space as \cite{michel_Introduction_2023}, 
\begin{equation}
    \frac{\delta \widetilde{n}_s(\mathbf{k})}{n_{e0}} = -\frac{m_ec^2}{2T_e}A_k\widetilde{|a_{3\omega}|^2}.
    \label{eq:Speckle-drivenPeruturbations}
\end{equation}
Here, $n_{e0}$ is the background electron density, $m_e$ is the mass of an electron, and $c$ is the speed of light. $A_k$ is a nonlocal transport coefficient, which must be considered due to the small scale length associated with the speckles produced by many overlapped beams \cite{bychenkov_Theory_2000}. When temperature gradient scale lengths are comparable to the electron mean free path, kinetic effects become important and lead to deviations from classical transport theory \cite{spitzer_Transport_1953, braginskii_Transport_1965}. The threshold for the nonlocal transport regime is given by $\text{Kn}> 0.06/\sqrt{Z}$ \cite{bychenkov_Nonlocal_1995}, where $\text{Kn} = \lambda_{mfp}/L_T$ is the thermal Knudsen number: the ratio of the electron mean free path, $\lambda_{mfp}$, and the temperature gradient scale length, $L_T = |T_e/\nabla T_e|$. The Knudsen number of the highest-intensity speckles is $\text{Kn}\sim10^1$-$10^2$, well above the nonlocal threshold. The temperature profile of these speckles is approximated by a spherical Gaussian \cite{senecha_temperature_1998} with a Gaussian waist of $\lambda_{3\omega}(f_\#)_{eff}\sim 351$ nm. The code uses the fit for $A_k$ from Ref. \cite{bychenkov_Theory_2000} that incorporates corrections to the ponderomotive force, inverse Bremsstrahlung heating, thermal conductivity, and ion convection \cite{bychenkov_Nonlocal_1995},
\begin{equation}
    A_k = \frac{1}{2} + Z\left( \frac{0.074}{\xi^2} + \frac{0.88}{\xi^{4/7}} + \frac{2.54}{1 + 5.5\xi^2}\right)
    \label{eq:NLCorrection}
\end{equation}
with $\xi = \sqrt{Z}k\lambda_{mfp}$. $k^2 = k_x^2 + k_y^2 +k_z^2$ is the fluctuation wavenumber. To accurately capture the 3-dimensional speckle distribution, the normalized vector potential of the overlapped heater beams, $a_{3\omega}(\mathbf{r})$, was calculated using the real beam geometry (cf. Fig. \ref{fig:setup}). Individual beams were modeled on local grids according to Ref. \cite{min-roberts_Statistical_2026} with a circular near-field aperture and random phases between 0 and 2$\pi$ distributed across the array. The fields of each beam were then Fresnel propagated \cite{goodman_Introduction_2017} to TCC in the far field and summed on a global grid.

The thermal density fluctuations arising from discrete particle effects were calculated using the spectral density $S(\mathbf{k},\omega)$ given in Min-Roberts \textit{et al.} \cite{min-roberts_Statistical_2026} (Eqs. (65) and (66)), based on the Maxwellian dynamic structure factor \cite{froula_Plasma_2010} evaluated at $\omega = 0$. %The spatial spectrum was then obtained by numerical integration over frequency.

Since the total heater-beam intensity contains pairwise products of the 14 individual beam fields, the speckle-driven density fluctuation spectrum from Eq. \ref{eq:Speckle-drivenPeruturbations} can be expressed using Wick's theorem for statistically independent Gaussian speckle fields as
\begin{equation}
    S_{\delta n_s \delta n_s}(\mathbf{k})\propto |A_k|^2 \sum_{m,n=1}^{14}|\hat{\mathbf{p}}_m\cdot\hat{\mathbf{p}}_n|S_{mn}(\mathbf{k}),
\end{equation}
where $\hat{\mathbf{p}}_m$ is the polarization of beam $m$ and $S_{mn}$ is the spectral contribution associated with the beam pair $(m,n)$ \cite{min-roberts_Statistical_unpublished}. The terms with $m=n$ describe speckle fluctuations within individual beams, while the terms with $m\neq n$ describe interference between crossing beams. 

\begin{figure}[t]
    \centering
    \includegraphics[trim = 10.7cm 0cm 10.7cm 7.3cm, clip, width=1.00\linewidth]{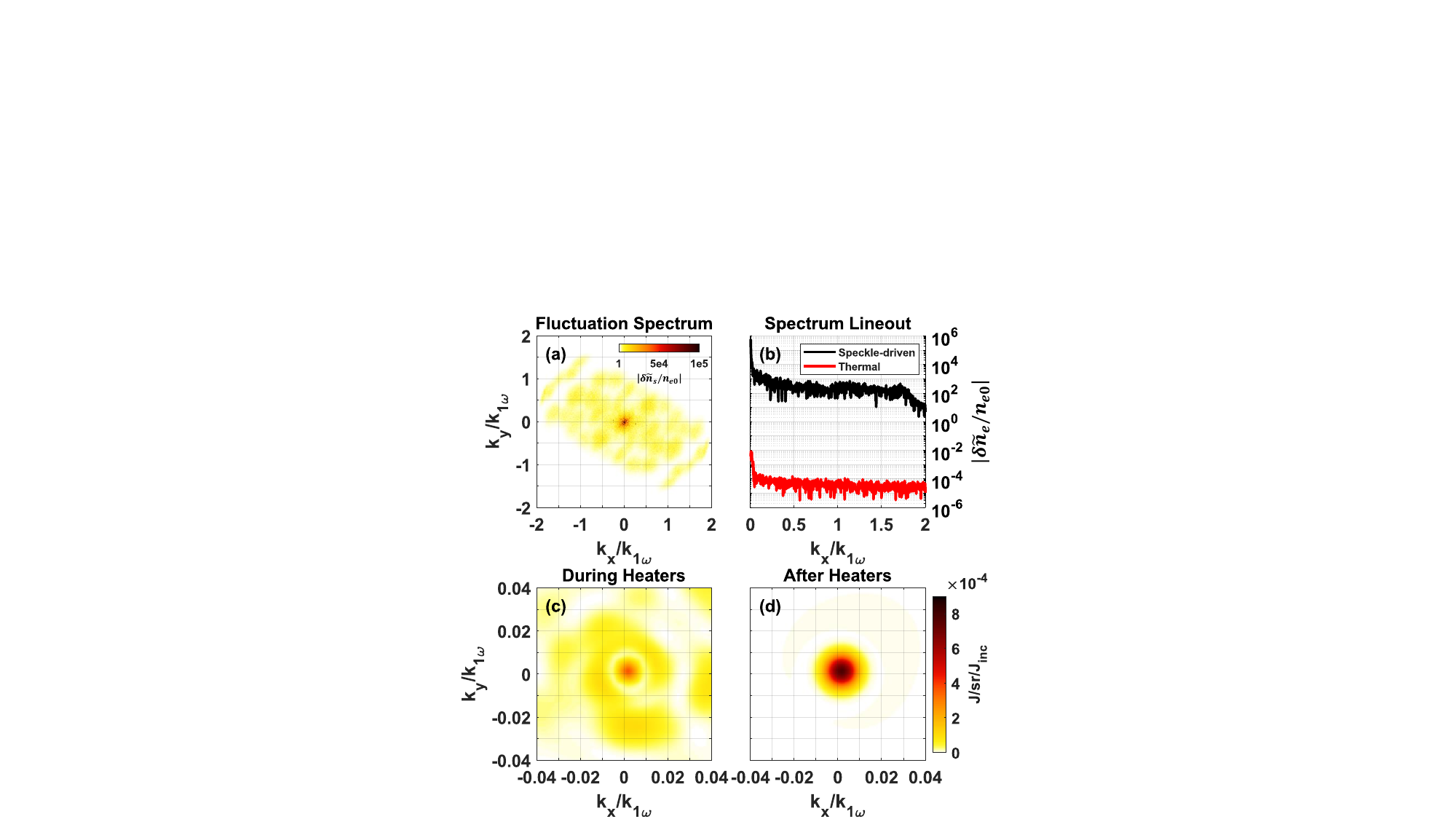}
    \caption{(a) Steady-state speckle-driven density fluctuation spectrum at $k_z = 0$. (b) Lineouts along $k_y = k_z = 0$ comparing the speckle-driven fluctuation spectrum (black) and the thermal fluctuation spectrum (red). Simulated beam spray from propagation through plasma with (c) and without (d) speckle-driven fluctuations.}
    \label{fig:DensityFluctuations}
\end{figure}

The speckle-driven density fluctuation spectrum, $|\delta \tilde{n}_s| \sim \sqrt{S_{\delta n_s \delta n_s}(\mathbf{k})}$, for the 14-beam configuration is shown in Fig. \ref{fig:DensityFluctuations}(a). The distinct elliptical features arise from interference between crossing heater beams, whereas the low-wavenumber continuum is dominated by contributions associated with individual-beam speckles. A lineout along $k_y=k_z=0$ in Fig. \ref{fig:DensityFluctuations}(b) compares the speckle-driven and thermal fluctuation spectra. The speckle-driven fluctuations are substantially larger than the thermal fluctuations at small wavenumbers, particularly in the range $k_\perp/k_{1\omega}\sim 0.02-0.04$. These transverse wavenumbers correspond to characteristic small-angle scattering of approximately 20--40 mrad, comparable to the measured $33.8$ mrad 95\%-encircled-energy angle with the heater beams on. This indicates that the low-wavenumber speckle continuum contributes to the observed angular broadening of the 1$\omega$ beam.

The real-space density perturbations corresponding to Figs. \ref{fig:DensityFluctuations}(a) and (b) had a volume-averaged amplitude of $\langle|\delta n_s/n_{e0}|\rangle = 0.18$ $\pm$ $0.11$. These simulations were run with an averaged overlapped heater beam intensity of $8\times10^{13}$ W/cm$^2$. For comparison, the stationary thermal fluctuations had a volume-averaged amplitude of $\langle|\delta n_t/n_{e0}|\rangle = 1.04 \pm0.79 \times 10^{-9}$.

Simulated beam spray results shown in Figs. \ref{fig:DensityFluctuations}(c) and (d) show that increased beam spray is a clear signature of the presence of speckle-driven density fluctuations. In these simulations, the 1$\omega$ beam was propagated using the paraxial wave equation
\begin{equation}
\nabla_\perp^2 a_{1\omega} + 2ik_{1\omega}\partial_z a_{1\omega} = \frac{\omega_{pe0}^2}{c^2}\frac{\delta n_e}{n_{e0}} a_{1\omega},
\end{equation}
where $\nabla_\perp^2 = \partial_x^2 + \partial_y^2$ is the transverse Laplacian, $a_{1\omega}(\mathbf{r}_\perp, z)$ is the normalized vector potential of the 1$\omega$ beam, and $\omega_{pe0}$ is the background plasma frequency. The 1$\omega$ beam propagates in the $z$-direction, such that $\mathbf{r}_\perp$ denotes the transverse coordinate.

The total density perturbations are taken as the sum of thermal, speckle-driven, and 1$\omega$ beam-driven density perturbations,
\begin{equation}
    \delta n_e(\mathbf{r}_\perp, z) =  \delta n_s(\mathbf{r}_\perp, z) + \delta n_t(\mathbf{r}_\perp, z) + \delta n_{1\omega}(\mathbf{r}_\perp, z).
    \label{eq:plasma-density fluctuations}
\end{equation}
The first two terms are described above, while the third term is the density perturbation due to the ponderomotive force of the 1$\omega$ beam, $\delta n_{1\omega}/n_{e0} = -(m_ec^2/4T_e)|a_{1\omega}|^2$. The steady-state model of 1$\omega$ beam-driven perturbations leads to self-focusing of the beam. To adjust simulations to match the short timescale in experiments, a defocusing phase was applied to the beam. The applied phase had the same radius of curvature but with the opposite sign as the wavefront curvature induced by the 1$\omega$ beam's steady-state density profile.

Simulations shown in Fig. \ref{fig:DensityFluctuations}(c) retain all three density terms, while simulations shown in (d) retain only the last two. The increased angular broadening in Fig. \ref{fig:DensityFluctuations}(c) demonstrates the contribution of the speckle-driven density fluctuations to the beam spray, consistent with local filamentation and self-focusing seeded by low-wavenumber fluctuations identified in Fig. \ref{fig:DensityFluctuations}(b). These simulations used an 800-$\mu$m-length plasma with uniform plasma conditions corresponding to plateau values in Fig. \ref{fig:TSData}(e) and a 1$\omega$ beam at an intensity of $1.4\times10^{15}$ W/cm$^2$.

In this Letter, we show that spatially smoothed laser beams drive plasma-density perturbations that induce beam spray in an unsmoothed beam. Experiments indicate that speckled beams sustain increased levels of long-wavelength steady-state fluctuations, which lead to small-angle beam spray. Modeling of the steady-state fluctuations shows amplitudes much larger than zero-frequency thermal fluctuations at an average overlapped intensity of $8\times10^{13}$ W/cm$^2$. Paraxial propagation simulations show these fluctuations induce local self-focusing and filamentation in the 1$\omega$ beam, leading to beam spray consistent with that seen in experiment.

\textit{Acknowledgements}---K.R.M. would like to thank M.A. VanDusen-Gross and M.V. Ambat for fruitful discussions, as well as the OMEGA 60 and OMEGA EP operations teams for their support on these experiments. 

This material is based upon the work supported by the Department of Energy [National Nuclear Security Administration] University of Rochester “National Inertial Confinement Fusion Program” under Award No. DE-NA0004144 and the Department of Energy, Office of Science, Office of Fusion Energy, under Award Numbers DE-SC0021057 and DE-SC0024863.

\appendix*
\section{Diagnostics}
Light that was Thomson scattered from the 2 J 2$\omega$ probe by plasma-density fluctuations was collected at an angle of 60$^\circ$ to the incident probe by the Thomson scattering collection system (TSC) \cite{katz_reflective_2013}. This light was transported to a spectrometer pair where the spatially resolved electron and ion-spectral features were recorded with spectral resolutions of 1.2 nm and 0.03 nm, respectively, and a spatial resolution of 20 $\mu$m. The measurement was time-integrated; the 100 ps probe set the effective integration time of the measurement.

The infrared transmitted beam diagnostic (IR-TBD), measured the time-integrated radiant intensity profile of the 1$\omega$ beam. The 1$\omega$ beam was intercepted by a transmissive diffuser and imaged to a charge-coupled device as part of the IR-TBD. The IR-TBD is based on the P9 transmitted beam diagnostic \cite{katz_transmitted-beam_2021} and the scattered-light uniformity imager \cite{edgell_scattered-light_2022}. The diffuser was positioned 1.1 m from TCC in the near field opposite the 1$\omega$ beam's final focusing optic. The diagnostic had an f/12.2 field of view and an angular resolution of 0.7 mrad.

\nocite{*}
% \section*{References}
\bibliography{bibliography}% Produces the bibliography via BibTeX.

\end{document}